\documentclass[a4paper,12pt]{article}
\usepackage[pctex32]{graphics}

\begin{document}
\topmargin 0pt \oddsidemargin 0mm
\newcommand{\beq}{\begin{equation}}
\newcommand{\eeq}{\end{equation}}
\newcommand{\beqa}{\begin{eqnarray}}
\newcommand{\eeqa}{\end{eqnarray}}
\newcommand{\sr}{\sqrt}
\newcommand{\fr}{\frac}
\newcommand{\mn}{\mu \nu}
\newcommand{\G}{\Gamma}

\begin{titlepage}

\vspace{5mm}
\begin{center}
{\Large \bf  Thermodynamics of  Schwarzschild-de Sitter black
hole: thermal stability of Nariai black hole} \vspace{12mm}

{\large   Yun Soo Myung \footnote{e-mail
 address: ysmyung@inje.ac.kr}}
 \\
\vspace{10mm} {\em Institute of Mathematical Science and School of
Computer Aided Science \\ Inje University, Gimhae 621-749, Korea}
\end{center}
\vspace{5mm} \centerline{{\bf{Abstract}}}
 \vspace{5mm}
We study thermodynamics of the  Schwarzschild-de Sitter black hole
 in five dimensions by introducing two
temperatures based on the standard and Bousso-Hawking
normalizations. We use  the first-law of thermodynamics to derive
thermodynamic quantities. The two temperatures indicate that the
Nariai black hole is thermodynamically unstable. However, it seems
that black hole thermodynamics favors the standard normalization,
and does not favor the Bousso-Hawking normalization.
\end{titlepage}
\newpage
\renewcommand{\thefootnote}{\arabic{footnote}}
\setcounter{footnote}{0} \setcounter{page}{2}

\section{Introduction}
 The Schwarzschild black hole with negative
specific heat is  in an unstable equilibrium with the  heat
reservoir of the temperature $T$~\cite{GPY}. Its fate under small
fluctuations will be either to decay to  hot flat space by Hawking
radiation or to grow without limit by absorbing thermal radiation
in the heat reservoir~\cite{York}. This means that an isolated
black hole is never in thermal equilibrium.  There exists a way to
achieve a stable black hole in an equilibrium with the heat
reservoir. A black hole could be rendered thermodynamically stable
by placing it in AdS space. An important point  to understand is
how a black hole with positive specific heat could emerge from
thermal radiation through a phase transition. To this end, the
Hawking-Page phase transition between thermal AdS space and
Schwarzschild-AdS black hole was introduced~\cite{HP,BCM,Witt}.

Furthermore,  a  thermodynamic  similarity between the event
horizon of a black hole  and the cosmological horizon of de Sitter
space has been established since the work of
Gibbons-Hawking~\cite{GH}. The key point is that a cosmological
horizon possesses temperature and entropy. Ginsparg and Perry have
studied the thermal properties of Schwarzschild-de Sitter black
hole (SdS)~\cite{GP}. However, an issue of  the negative mass to
the cosmological horizon has appeared when using the first-law of
thermodynamics to derive thermodynamic quantities~\cite{GH,CG}.
This problem arises because the surface gravity $\kappa_{C}$ of
the cosmological horizon is negative. Using the first law of
$dM_{C}=[\kappa_{C}/8\pi]dA$ leads to the mentioned result. A way
to resolve this issue is to calculate the  mass of cosmological
horizon using the Brown-York approach in the asymptotic
future~\cite{TDS}.

 It is known that the SdS  is
curious but a difficult object to analyze its thermodynamic
properties because a black hole is inside the fixed cosmological
horizon. The cosmological horizon may play a role of the heat
reservoir for a black hole  like the AdS space. In order to
investigate the SdS, one introduces two kinds of temperature based
on the standard and Bousso-Hawking normalizations. The standard
normalization provides the Hawking
  temperature $T^{E}_H$ and Gibbons-Hawking temperature $T^{C}_H$
   for event  and cosmological horizons, respectively~\cite{myungsds}. They
behave differently but have  the  zero temperature at the Nariai
case which corresponds to the maximum black hole and minimum de
Sitter space. These temperatures were derived  by the analogy with
asymptotically flat and AdS space. In the case of asymptotically
flat spacetimes, a standard  method to obtain the surface gravity
is to choose the Killing field that goes to a unit
time-translation at infinity.  An observer to stay there does not
feel any acceleration. However, there is no asymptotic region and
thus no preferred observer in de Sitter spacetimes. Hence one has
to introduce another normalization to define appropriate
temperatures. This is the Bousso-Hawking normalization~\cite{BH1}.
At the point $r=r_0$ where the metric function satisfies
$h'(r_0)=0$, the black hole attraction and the cosmological
repulsion exactly cancel out and thus one may achieve the zero
acceleration inside the cosmological horizon. Including this
normalization into the expression of the surface gravity, one
finds the Bousso-Hawking temperatures $T^{E/C}_{BH}$. These do not
vanish in the Nariai limit but approach a constant
value~\cite{BH2,CG,KGB}. However, one has to realize that the
temperature $T^C_{BH}$ of cosmological horizon is just an
extension of $T^E_{BH}$  of the event horizon and thus an
important property of the degenerate horizon at $r=r_0$ may be
lost  for the thermodynamic purpose.

In this work, we investigate thermal properties and  phase
transition of the SdS by introducing two temperatures. Especially,
we reexamine the thermal stability of the Nariai black hole which
was considered in Ref.\cite{myungsds}.

Our study is based on the on-shell observations of temperature,
heat capacity and  free energy as well as the off-shell
observations of  generalized  free energy, deficit angle and
$\beta$-function.  In general, the on-shell thermodynamics implies
equilibrium thermodynamics and thus the first-law of
thermodynamics holds for this case. Hence it describes
relationships among thermal equilibria, but not the transitions
between equilibria. On the other hand, the off-shell
thermodynamics is designed for the description of off-equilibrium
configurations~\cite{FFZ,off}.  This is suitable for the
description of transitions between thermal equilibria. We note
that the first-law of thermodynamics does not hold for off-shell
thermodynamics. We believe that the thermodynamic study on the SdS
is very helpful to understand de Sitter spacetimes~\cite{HKL}
because other analysis of perturbations under the SdS background
~\cite{KOY} is more restrictive than thermodynamic analysis in de
Sitter spacetimes.

\section{The standard normalization}

We wish  to study  the thermal property of a black hole in de
Sitter space. For this purpose,  we consider the Schwarzschild-de
Sitter black hole in five-dimensional spacetime~\cite{CAI2} \beq
ds^{2}_{SdS}= -h(r)dt^2 +\fr{1}{h(r)}dr^2 +r^2 d\Omega_3^2,
\label{SDS} \eeq where the metric function $h(r)$ is given by \beq
h(r)=1-\fr{m}{r^2}- \fr{ r^2}{\ell^2}.\eeq Here $m$ is a reduced
mass of the black hole and $\ell$ is the curvature radius of de
Sitter spacetime. In the case of $m=0$ (no black hole), we have
the pure de Sitter space with the largest cosmological horizon
($r_C=\ell$).  $m \not=0$ generates the SdS black hole. From the
condition of $h(r_{C/E})=0$, one finds that the cosmological and
event horizons are located at \beq \label{3EH} r_{C/E}^2=
\fr{\ell^2}{2}\Big(1\pm \sqrt{1 - m/m_0}\Big) \eeq with
$m_0=\ell^2/4=r_0^2/2$. We classify three cases with
$r_0=\ell/\sqrt{2}$: $m=m_0(r=r_0)$, $m>m_0$, and $m<m_0$. The
case of $m=m_0$ corresponds to the maximum black hole with the
minimum cosmological horizon, the Nariai black hole. Here we have
the degenerate horizon of $r_0=r_{E}=r_{C}$. A large black hole of
$m>m_0$ is not allowed in de Sitter space. The case of $m<m_0$
corresponds to a small black hole inside the cosmological horizon.
In this case a cosmological horizon is located  at
 $r_{C} \simeq \ell\sqrt{1-m/4m_0}$, while  an event
horizon is at  $r_{E} \simeq \sqrt{m}$. Hence, restrictions on
$r_E$ and $r_C$ are given by \beq \label{2Ineq} 0< r_{E} \le
r_0,~~ r_0 \le r_{C} <\ell. \eeq One  expects that as a reduced
mass $m$ approaches the maximum value of $m=m_0$, the small black
hole increases to the Nariai black hole at $r_E=r_0$
 by absorbing radiation ($\overrightarrow{EH}$), whereas the
cosmological horizon decreases to the minimum value of $r_C=r_0$
 by emitting radiation ($\overleftarrow{CH}$). This was the
Hawking-Page transition (HP) for obtaining a large, stable black
hole in de Sitter space~\cite{myungsds}. Also we note that the
size of black hole is closely related to the size of cosmological
horizon.

The energy and entropy for two horizons take the forms
\begin{equation}
E_{E/C} =\pm \Big(\fr{3V_3 m_{E/C}}{16 \pi G_5} - E_0\Big)~{\rm
with}~m_{E/C}=r_{E/C}^2 \Big(1-\fr{r_{E/C}^2}{\ell^2}
\Big),~S_{E/C}=\fr{V_3r_{E/C}^3}{4G_5},
\end{equation}
where $V_3$ is the volume of a unit three-dimensional sphere
$\Omega_3$ and $E_0=\fr{3V_3 m_{0}}{16 \pi G_5}$ is the energy of
the Nariai black hole. We note here that $E_{E}\le 0$, while
$E_{C}\ge 0$.  In this case, there is no energy gap between two
horizons $(E_E=E_C)$ at $r_+=r_0$. We use these definitions of
energy since the fixed-$\ell$ ensemble of de Sitter space is
similar to the fixed-charge $Q$ ensemble in the
Reissner-Norstr\"om-AdS black holes~\cite{CEJM}. However, the
energy was used without $E_0$ in
Refs.\cite{CM,NOO,myungcqg,myungsds}.
 The thermodynamic quantities of temperature, heat capacity, and free energy for the two horizons   are given
by \beqa
\label{2TQ}T_H^{E/C}&=&\mp\Big(\frac{2r_{E/C}^2-\ell^2}{2\pi
\ell^2 r_{E/C}}\Big),~ C_{E/C}= 3
\fr{2r_{E/C}^2-\ell^2}{2r_{E/C}^2+\ell^2}S_{E/C},\\
F_{E/C}&=&\pm\fr{V_3 r_{E/C}^2}{16 \pi
G_5}\Big(\fr{r_{E/C}^2}{\ell^2}+1 \Big)\mp E_0. \eeqa
 Hereafter we use the
normalization of $V_3/16 \pi G_5=1$ for simplicity. It is easily
checked that the first-law of thermodynamics holds for two
horizons, \beq dE_{E/C}=T^{E/C}_HdS_{E/C}. \eeq

Imposing the equilibrium condition $T=T_H$, we obtain a small,
unstable black hole of  size \beq \label{ubhs} r_u =\frac{\pi
\ell^2T}{2}\Bigg[-1+\sqrt{1+\frac{8}{(2\pi \ell T)^2}}\Bigg]\eeq
and a large, stable cosmological horizon of  size \beq
\label{sbhs} r_s=\frac{\pi
\ell^2T}{2}\Bigg[1+\sqrt{1+\frac{8}{(2\pi \ell T)^2}}\Bigg].\eeq
\begin{figure}[t!]
 \centering
   \includegraphics{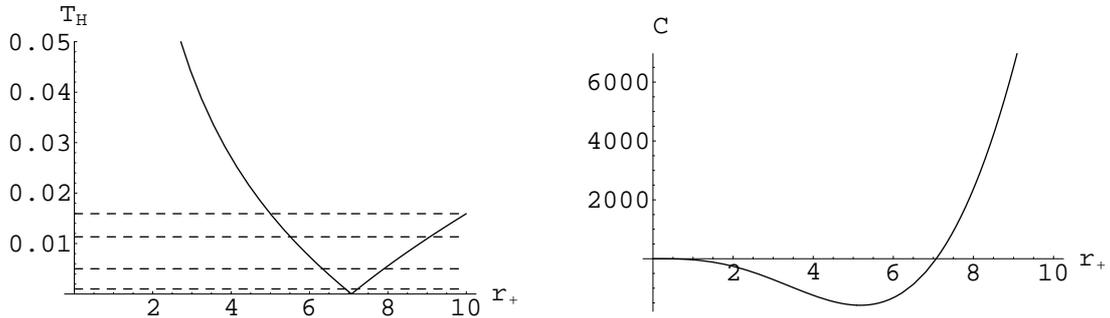}
    \caption{Temperature and heat capacity for SdS with $\ell=10$ and $r_0=7.07$.
     Here $r_+=r_E~(r_C)$ are confined to $0<r_E \le r_0~(r_0 \le
    r_C <\ell)$.
     At the left graph, the solid curve represents the temperature of the event horizon $T^{E/C}_H$,
     while the dashed lines denote  four external temperatures of $T=T^C_{max}(=0.016),0.011,0.005,0.001$
      from top to bottom.
     } \label{fig1}
\end{figure}

\begin{figure}[t!]
 \centering
   \includegraphics{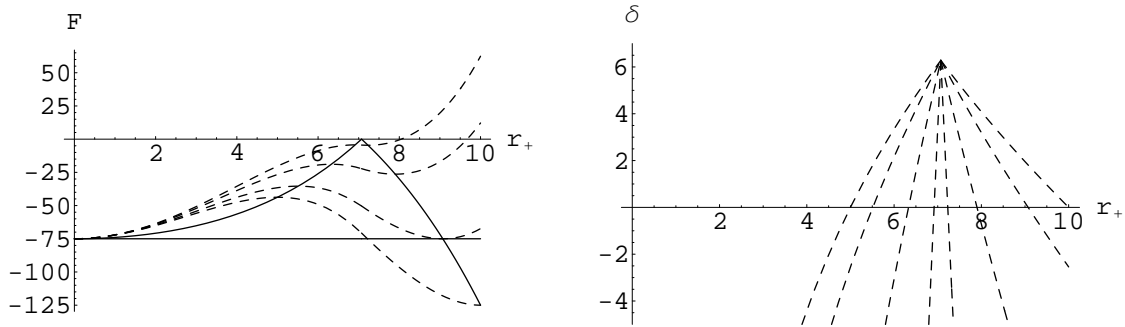}
    \caption{The graphs of free energy and  deficit angle $\delta_{E/C}$ for  SdS.
    Here $r_+$ represents $r_E$ for the event horizon and $r_C$
    for the cosmological horizon. At the left graph, the solid curve represents the free energy $F_{E/C}$,
     while the dashed lines denote off-shell free energy  $F^{off}_{E/C}(r_+,T)$
     for    temperatures of $T=0.001,0.005,T_c(=0.011),0.016$ from top to bottom.
     The   reverse order of $T$ is for the deficit angle $\delta_{E/C}$. } \label{fig2}
\end{figure}
As is shown in Fig. 1, the temperatures $T^{E/C}_H$ behave
differently. We find  two thermal equilibria for the range of
$0\le T \le T^C_{max}=1/2 \pi \ell$. For $T>T^C_{max}$, there
exists  one unstable equilibrium. Four temperatures $\{T\}$ are
introduced to investigate the phase transition. For these
temperatures, we have unstable equilibria of
$\{r_u\}=\{5,5.51,6.33,6.91\}$  and stable equilibria of
$\{r_s\}=\{10,9.07,7.90,7.23\}$. Even though  the temperature
graph shows the key  property, we need to introduce other
quantities for a complete analysis of thermodynamic stability and
phase transition.

 From the graph of heat capacity, we find that the event
horizon $r_E$ is locally unstable because of negative heat
capacity, whereas the cosmological horizon $r_C$ is locally stable
because of positive heat capacity. A global stability of black
hole is achieved only when $C>0$ and $F<0$. The cosmological
horizon of $r_C>r_0$ seems to be globally stable, as is shown in
Fig. 2. However, such thermodynamic arguments describe
relationships among thermal equilibria, but  not the transitions
between equilibria. In order to describe  transitions between
thermal equilibria, we need to introduce the off-shell free
energy, deficit angle, and off-shell $\beta$-function
as~\cite{myungjmpa}

\beqa
F_{E/C}^{off}(r_+,T)&=&E_{E/C}-TS_{E/C},~\delta_{E/C}(r_+,T)=2\pi
\Big(1-\frac{T^{E/C}_H}{T}\Big),\\
\beta_{E/C}(r_+,T)&=&-6r_+^2\delta_{E/C}(r_+,T). \eeqa

We use the off-shell free energy to study the growth of a black
hole~\cite{FFZ}. In order to investigate the off-shell process
explicitly,  we consider  the deficit angle $\delta_{E/C}$.  The
range of deficit angle is $0\le \delta_{E/C} \le 2\pi$ for the
proper transition between two black holes. $\delta_{E/C}$ has the
maximum value of $2\pi$ at the extremal point and it is zero at
the equilibrium point of $T=T_H$. This implies that the Nariai
configuration at $r_+=r_0$ has the narrowest cone of the shape
($\prec$) near the horizon, while the geometry at $T=T_H$ is a
contractible manifold ($\subset$) without conical singularity. For
any off-shell process of  the growth of black hole, we  have
$0<\delta_{E/C}<2\pi$ and a conical singularity of the shape ($<$)
is allowed near the horizon~\cite{FFZ,off,myungjmpa}.
 Also, the off-shell $\beta$-function is
introduced to measure the mass of a conical singularity at the
event horizon~\cite{off}. Hereafter, we do not consider the
$\beta$-function because it is proportional to the deficit angle
$\delta_{E/C}$.

All equilibria of $\{r_u\}$ and $\{r_s\}$ could be reproduced  by
each  condition of $F^{off}_{E/C}=F_{E/C}$ and $\delta_{E/C}=0$.
 We know that the black hole  is quite different
from the cosmological horizon because the former is unstable,
while the latter is stable. The HP may occur for $T>T_c$ where
$T_c=0.011$ is determined from the equilibrium condition of
$F^{off}_{C}(r_c,T_c)=F_C(r_c)=F_E(0)$ at $r_c=9.07$ with
$F_E(0)=-75$. We note a sequence of free energy of
$F_C(\ell)<F_E(0)<F_{E/C}(r_0)$, which means  that the pure de
Sitter space at $r_+=\ell$ is globally stable and the Nariai black
hole at $r_+=r_0$ is unstable.  As $T \to 0$, $F^{off}(r_+,T)$
connects the point of $r_+=0$ to the Nariai case. On the other
hand, as $T \to T^C_{max}$, $F^{off}(r_+,T)$ connects the point of
$r_+=0$ to the pure de Sitter space through the unstable black
hole at $r_+=r_u$.
 For $T>T_c$,
the pure de Sitter space  is more favorable than the Nariai case,
while for $T<T_c$, the pure de Sitter space is less favorable than
the Nariai case. This implies that the HP of
$\overrightarrow{EH}\overleftarrow{CH}$ is unlikely to occur by
absorbing radiation, while the evaporating process of
$\overleftarrow{EH}\overrightarrow{CH}$ is likely to occur by
emitting radiation.

We note that  for $T<T_C$, $F^{off}_E(r_u,T)$ and
$F^{off}_C(r_s,T)$  are greater than $F_E(0)$. On the other hand,
for $T>T_C$, $F_E(0)$ is between $F^{off}_E(r_u,T)$ and
$F^{off}_C(r_s,T)$.  There exists an evaporating process from
 $r_E=r_0$ to $r_E=0$ even for $T\simeq 0$.
 This
shows  that the Nariai black hole is not a globally stable object,
whereas the pure de Sitter space  is a globally stable object. The
shapes of free energy and its off-shell free energy are similar to
those for the Schwarzschild-AdS black hole. All of deficit angles
$\delta_{E/C}$  are positive for  proper transitions between $r_u$
and $r_s$. The differences are the downward shift of free energy
and the peak point  at $r_+=r_0$ as the extremal point.  Hence,
the  HP of $\overrightarrow{EH}\overleftarrow{CH}$ may be excluded
from the candidate for  phase transition of  the SdS. This is an
opposite conclusion to the previous result based on the
discontinuous free energy~\cite{myungsds}.

\section{The Bousso-Hawking normalization}

The new temperatures based on the Bousso-Hawking normalization
take the form~\cite{BH1,BH2,CG,KGB}
\begin{equation} \label{bht}
T^{E/C}_{BH}=\frac{T_H^{E/C}}{
\sqrt{h(r_0)}}=\mp\frac{1}{\sqrt{1-\frac{2r_+\sqrt{\ell^2-r_+^2}}{\ell^2}}}
\Big(\frac{2r_+^2-\ell^2}{2\pi \ell^2 r_+}\Big),
\end{equation}
where $r_+=r_E(r_C)$ go with $0 \le r_E \le r_0(r_0 \le r_C \le
\ell)$. Here we check that in the Nariai limit, the Bousso-Hawking
temperatures for event and cosmological horizons approach a
constant value as
\begin{equation}
\lim_{r_+ \to r_0}T^{E/C}_{BH}=\frac{1}{\pi \ell}.
\end{equation}
Here   one takes  the limit from smaller value for computing
$T^{E}_{BH}(r_0)$, while for  $T^{C}_{BH}(r_0)$, one takes the
limit from larger value.

 Assuming that the first-law of thermodynamics
\begin{equation}
d\tilde{E}_{E/C}=T^{E/C}_{BH}dS_{E/C}
\end{equation}
 holds for this
normalization, we obtain  the corresponding energy from its
integration over $r_+$ as
\begin{equation}
\tilde{E}_{E/C}=\mp \frac{2}{\ell} \Bigg(\ell^2
+r_+\sqrt{\ell^2-r_+^2}\Bigg)\sqrt{\ell^2-2r_+\sqrt{\ell^2-r_+^2}}.
\end{equation}
The heat capacity is defined to be
\begin{equation}
\tilde{C}_{E/C}=\frac{d\tilde{E}_{E/C}}{dT^{E/C}_{BH}}=\frac{r_+(2r_+^2-\ell^2)\sqrt{\ell^2-r_+^2}
(\ell^2-2r_+\sqrt{\ell^2-r_+^2})}
{\ell^2\Big(2r_+^3-3\ell^2r_++(2r_+^2+\ell^2)\sqrt{\ell^2-r_+^2}\Big)}.
\end{equation}
The on-shell free energy is defined by
\begin{equation}
\tilde{F}_{E/C}=\tilde{E}_{E/C}-T^{E/C}_{BH}S_{E/C}.
\end{equation}
On the other hand, the off-shell free energy is
\begin{equation}
\tilde{F}_{E/C}^{off}=\tilde{E}_{E/C}-TS_{E/C}
\end{equation}
The equilibrium condition of $d\tilde{F}^{off}_{E/C}/dr_+=0$
provides $T=T_{BH}^{E/C}$, which shows in turn  that
$\tilde{F}_{E/C}^{off}= \tilde{F}_{E/C}$. Similarly, we could
define the deficit angle $\tilde{\delta}_{E/C}(r_+,T)$ using the
temperatures $T^{E/C}_{BH}$.

\begin{figure}[t!]
 \centering
   \includegraphics{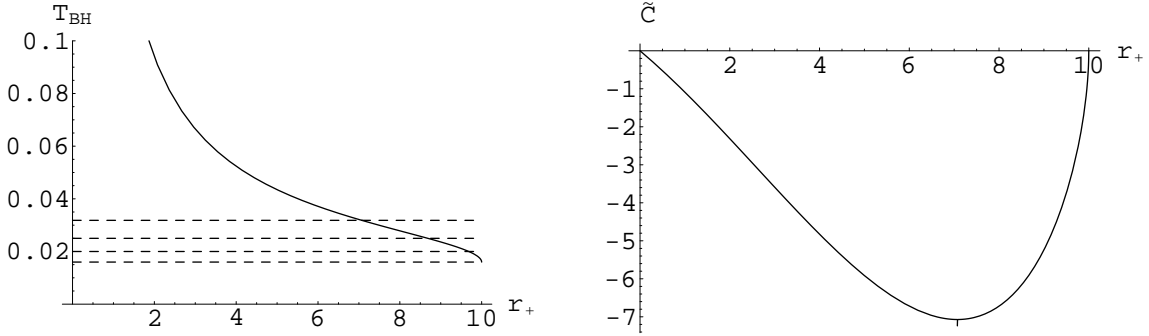}
    \caption{Temperature,  and heat capacity $\tilde{C}_{E/C}$ for SdS with $\ell=10$
    and $r_0=7.07$.
     Here $r_+=r_E~(r_C)$ are confined to $0<r_E \le r_0~(r_0 \le
    r_C <\ell)$.
     At the left graph, the solid curve represents the temperature of  $T^{E/C}_{BH}$,
     while the dashed lines denote  the external temperatures of $T=0.032,0.025,0.02,0.016$ from top to bottom.} \label{fig3}
\end{figure}

\begin{figure}[t!]
 \centering
   \includegraphics{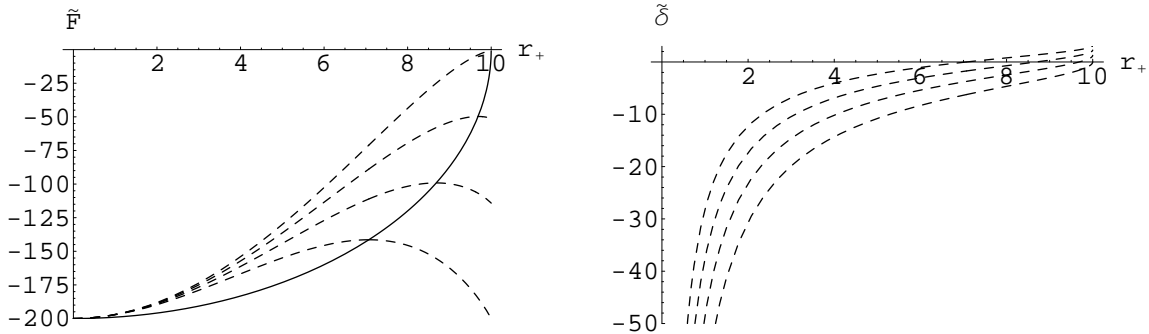}
    \caption{ The graph of free energy and  deficit angle $\tilde{\delta}_{E/C}$  for SdS with
    $T^{E/C}_{BH}$.
    Here $r_+$ represents $r_E$ for the event horizon and $r_C$
    for the cosmological horizon. At the left graph, the solid curve represents the free energy $\tilde{F}_{E/C}$,
     while the dashed lines denote  $\tilde{F}^{off}_{E/C}(r_+,T)$
     for  external temperatures of $T=0.016,0.02,0.025,0.032$ from top to bottom.
     The  reverse order of $T$ is for the deficit angle $\tilde{\delta}_{E/C}$. } \label{fig4}
\end{figure}

Now we are in a position to  discuss the thermal properties of the
SdS which are based on $T^{E/C}_{BH}$. From Fig. 3, it turns out
that the cosmological horizon branch is just an extension of the
event horizon branch. We have the temperature bound of  $T^E_{BH}
\ge T^C_{BH}$, where the equality holds for the Nariai black hole.
Here we have the range of temperature for the cosmological
horizon: $T^C_{BH}(\ell=10) \le T \le T^C_{BH}(r_0=7.07)$. Both
horizons are thermodynamically unstable because of
$\tilde{C}_{E/C}<0$.  We have $\tilde{C}_C(\ell)=0$ for the pure
de Sitter case and $\tilde{C}_E(0)=0$ for no black hole. From of
the free energy graph in Fig. 4, it follows that any Hawking-Page
phase transition would not occur between two branches because
thermal equilibria of the cosmological horizon are unstable
points. We observe that the deficit angle $\tilde{\delta}_{E/C}$
are positive  only for $r_u<r_+<\ell$, where unstable equilibria
$\{r_u\}=\{9.98,9.68,8.68,7.07\}$ are determined by the condition
of $T=T^C_{BH}$. Actually, this region is beyond thermal
equilibria. We point out that the Nariai back hole at $r_+=r_0$ is
nothing special in thermodynamic aspect.

 Finally, we  mention that the  Bousso-Hawking normalization does
 not provide attractive features for the thermodynamics of  SdS, because
 it does not make a significant
 distinction between the event and cosmological horizons.

\section{Discussion}
We start to discuss two limiting cases: a very small black hole
($m \ll m_0$) and a nearly degenerate Schwarzschild-de Sitter case
($m \simeq m_0$).

 For the
first case, the effect of the radiation coming from the
cosmological horizon is negligible and one would expect the
evaporating process to be similar to that of Schwarzschild black
hole. Thus we expect to have  the pure de Sitter space (no black
hole) as the stable ending point.

The second case corresponds to the near-horizon thermodynamics of
degenerate horizon~\cite{CFNN}. In  case of the  Nariai black
hole, the two horizons have the same size and the same
temperature. Hence they will be in thermal equilibrium. If one
considers a  perturbation of the geometry to cause the black hole
to become hotter than the cosmological horizon,  the thermal
condition of the Nariai black hole becomes unstable. Actually, the
thermal stability will be  determined by the sign of heat
capacity.

At this stage,  we would like to mention the Nariai phase
transition of the SdS at $T=0$. A previous work has shown that the
location $r_+=r_0$ is not only the critical point of phase
transition  but also the position of the stable cosmological
horizon. This arises because  an inappropriate form of the
discontinuous free energy was used to analyze the Nariai
configuration~\cite{myungsds}. In this work, we showed that  the
Nariai black hole is not a globally stable object when using the
continuous free energy. Instead, the pure de Sitter space plays
the role of a globally stable object. Consequently, the  HP of
$\overrightarrow{EH}\overleftarrow{CH}$ is unlikely to occur by
absorbing radiation from the cosmological horizon, while the
evaporating process of $\overleftarrow{EH}\overrightarrow{CH}$ is
likely to occur by emitting radiation. This is consistent with
intuitive thermodynamic arguments on the black hole in de Sitter
space.

If one uses the Bousso-Hawking temperatures, the Nariai black hole
is thermodynamically unstable because of their negative heat
capacity. Furthermore, it seems inappropriate to describe either
the Hawking-Page phase transition or the evaporation process by
using these temperatures.

At this stage, we would like to comment on another temperature
$\bar{T} \propto \sqrt{-h''(r_0)}$ of the SdS ~\cite{CN}. This
temperature is valid for the near-horizon region only because the
condition of $h''(r_0)\not=0$ implies the near-horizon  of the
degenerate horizon. For the whole region, it would be  better to
use the temperature (4.9) in Ref.\cite{CG} for four dimensions or
$T^{E/C}_{BH}$ in Eq.(\ref{bht}) for five dimensions.

In conclusion, it turns out that the Nariai black hole of
$r_E=r_0$ is a thermodynamically unstable object. The Hawking-Page
phase transition from $r_E=0$ to $r_E=r_0$ is unlikely to occur,
while the evaporation process from $r_E=r_0(=r_C)$ to
$r_E=0(r_C=\ell)$ is likely to occur when using the Hawking and
Gibbons-Hawking temperatures based on the standard normalization.

However, a  small group of works~\cite{HKL,NO,myungsds} supports
the stable Nariai black hole, while a large group of
works~\cite{BH2,DL,Bousso,Liu} shows the instability of Nariai
black hole. The former has used the standard normalization to
support the stability of the Nariai black hole, the latter has
used different schemes to show the instability. Here, we focus on
the thermodynamic stability of the Nariai black hole.  A black
hole is thermodynamically unstable when its heat capacity is
negative ($C<0$). Furthermore, a global stability of black hole is
achieved only when $C>0$ and $F<0$. As is shown Fig. 1, we have
the zero heat capacity for the case of the Nariai black hole in
the standard normalization, which  means that the issue of thermal
stability remains unclear and it should be further resolved  by
choosing an appropriate free energy. We used the discontinuous
free energy in Ref.\cite{myungsds}, where the Nariai black hole
was shown to be stable. In this work, we used an appropriate  free
energy to show the unstable Nariai black hole. On the other hand,
we find from Fig. 3 that the heat capacity of the Nariai black
hole is always negative. This means that the Nariai black hole is
unstable when using the Bousso-Hawking normalization.

Finally, this work supports the instability of  Nariai black hole.

\section*{Acknowledgments}
The author thanks Jin-Ho Cho  for helpful discussions. This work
was supported by the Korea Research Foundation
(KRF-2006-311-C00249) funded by the Korea Government (MOEHRD).

\end{document}